\documentclass[aps,prd]{revtex4-2}
\usepackage[utf8]{inputenc}
\usepackage[T1]{fontenc}
\usepackage{amsmath,amsthm}
\DeclareMathOperator{\tr}{Tr}
\DeclareMathOperator{\im}{Im}
\DeclareMathOperator{\diag}{Diag}
\newcommand{\me}{\mathrm{e}}
\newtheorem{thm}{Theorem}

\begin{document}
\title{\textbf{CP violation for four generations of quarks}}
\author{Ubaldo Cavazos Olivas}
\affiliation{National Centre for Nuclear Research,
Pasteura 7, 02-093 Warsaw, Poland}

\author{S. Rebeca Juárez Wysozka}\thanks{Supported in part by Proyecto SIP: 20195244 , Secretaría de Investigación y Posgrado, Beca EDI y Comisión de Operación y Fomento de Actividades Académicas (COFAA) del Instituto Politécnico Nacional (IPN), México.}
\affiliation{Departamento de Física, Escuela Superior de
    Física y Matemáticas, Instituto Politécnico
    Nacional, U.P. ``Adolfo López Mateos'',
    C.P.~07738 Ciudad de México, Mexico}
  
\author{Piotr Kielanowski}
\affiliation{Departamento de F\'{i}sica, Centro de Investigaci\'on y de Estudios Avanzados, C.P. 07000 Ciudad de M\'exico, Mexico}

\begin{abstract}
  We discuss the generalization of the Jarlskog condition of CP
  conservation for the case of the Standard Model with 4~quark
  generations. We express this condition in terms of the
  3~\textit{Jarlskog} invariants of the CKM matrix. Next we present
  the test for the existence of the 4-th quark generation in terms of
  the \textit{Jarlskog} invariants involving only known particles.
\end{abstract}
\maketitle

\section{Introduction}\label{sec:intro}
The Standard
Model~\cite{Weinberg:1967tq,Glashow:1961tr,Salam:Nobel,Higgs:1964ia,Higgs:1964pj,Higgs:1966ev,Englert:1964et,tHooft:1971akt,tHooft:1971qjg}
(SM) is constructed in such a way that it reproduces all known
phenomenological information about the spectrum and interactions of
elementary particles. When one considers an extension of the SM, then
certain properties of the construction are automatically generalized,
but frequently generalizations are not simple and require a careful
analysis.

Recently, a new evidence for a fourth, sterile neutrino, has
appeared~\cite{Aguilar-Arevalo:2018gpe}. This has again opened a
possibility for a fourth
generation~\cite{Jarlskog:1988ii,Frampton:1999xi} of quarks and
leptons in the SM, though its nature, if it exists, might not be clear
and it may also require an extension of the Higgs
sector~\cite{Lenz:2013iha}. It should be also noted that the Standard
Model with four generations provides sufficient CP~violation to
explain the Universe baryon asymmetry~\cite{Hou:2008xd}.

In this paper we will concentrate our attention on the conditions for
the CP-conservation in the four generations SM (SM4). The general
(necessary and sufficient) condition for the CP conservation has been
formulated in~\cite{Bernabeu:1986fc,Gronau:1986xb} and requires that there exists a unitary
matrix $U_{L}$ such that the quark mass matrices $M_{q}$ fulfill the
following conditions
\begin{equation}
  \label{eq:1}
  U_{L}^{\dagger}H_{q}U_{L}=H_{q}^{*},\quad H_{q}=M_{q}M_{q}^{\dagger},\quad q=u,d.
\end{equation}
Condition~\eqref{eq:1} is valid for any number of generations, but it
is not expressed in terms of the observables. In Ref.~\cite{Bernabeu:1986fc,Gronau:1986xb}
there is a theorem that determines the equivalence of
condition~\eqref{eq:1} with the vanishing of the imaginary part of the
products of powers of the $H_{q}$ matrices, which are observables
(quark masses and elements of the Cabibbo-Kobayashi-Maskawa (CKM)
matrix). For the SM with 3~generations (SM3) condition~\eqref{eq:1} is
equivalent to vanishing of the following expression
\begin{multline}
  \label{eq:2}
  \im(\tr(H_{u}^{2}H_{d}H_{u}H_{d}^{2}))=(m_{t}^{2}-m_{c}^{2})(m_{t}^{2}-m_{u}^{2})
  (m_{c}^{2}-m_{u}^{2})\\ \times
  (m_{b}^{2}-m_{s}^{2})(m_{b}^{2}-m_{d}^{2}) (m_{s}^{2}-m_{d}^{2})
  \im(V_{ud} V_{cd}^{*}V_{cs}V_{us}^{*})=0.
\end{multline}
Eq.~\eqref{eq:2} is the necessary and sufficient condition for the CP
conservation in the SM with 3~generations. This condition states that
there is no CP violation if the \textit{Jarlskog} invariant
$\im(V_{ud} V_{cd}^{*} V_{cs} V_{us}^{*} )$~\cite{Jarlskog:1985ht,Jarlskog:9971505606} vanishes or
if any two masses within up or down quark sectors are equal. The
equality of masses within the up or down sectors implies new
properties of the CKM matrix: if one pair of quark masses within a
multiplet were equal (e.g., $m_{u}=m_{c}$) then the CKM matrix would
depend on two angles only (no CP violation) and if all quark masses of
the same type were equal (e.g., $m_{u}=m_{c}=m_{t}$), then the CKM
matrix would be an identity matrix. Such is the mechanism of
CP~conservation for the case of equal masses, so the necessary and
sufficient condition of CP~conservation for 3~generations in the SM3
model is only the vanishing of the \textit{Jarlskog} invariant. What
should be stressed, is that for the case of two equal quark masses the
CP~conservation requires that the equality of masses must be exact and
this would require some kind of \textit{fine tuning} and would
probably be a consequence of new conservation laws.

On the other hand the conditions for CP~conservation for the SM4, in
terms of observables, are more involved. It turns out that the
necessary and sufficient conditions for CP~conservation in case of
non-degenerate masses consist in the vanishing of the following
expressions~\cite{Bernabeu:1986fc,Gronau:1986xb}
\begin{equation}
  \label{eq:3}
  \begin{aligned}
    & \im(\tr(H_{u}^{2}H_{d}H_{u}H_{d}^{2}))=0,\quad
    \im(\tr(H_{u}^{2}H_{d}H_{u}H_{d}^{3}))=0\\
    & \im(\tr(H_{u}^{2}H_{d}^{2}H_{u}H_{d}^{3}))=0,\quad
    \im(\tr(H_{u}H_{d}H_{u}H_{d}^{2}H_{u}H_{d}^{3}))=0\\
    & \im(\tr(H_{u}^{3}H_{d}H_{u}H_{d}^{2}))=0,\quad
    \im(\tr(H_{u}^{3}H_{d}H_{u}H_{d}^{3}))=0.
  \end{aligned}
\end{equation}
Another set of conditions for the CP~conservation for the
4~generations SM was given in Ref~\cite{delAguila:1996pa} and reads
\begin{equation}
  \begin{aligned}
    & I_{1}= \im(\tr(H_{u}^{2}H_{d}H_{u}H_{d}^{2}))=0\\
    & I_{2}= \im(\tr(H_{u}^{3}H_{d}H_{u}H_{d}^{2}))=0\\
    & I_{3}= \im(\tr(H_{u}^{4}H_{d}H_{u}H_{d}^{2}-H_{u}^{3}H_{d}H_{u}^{2}H_{d}^{2}))=0\\
    & I_{4}= \im(\tr(H_{u}^{5}H_{d}H_{u}H_{d}^{2} -H_{u}^{4}H_{d}H_{u}^{2}H_{d}^{2} +H_{u}^{3}H_{d}H_{u}^{2}H_{d}H_{u}H_{d}))=0\\
    & I_{5}= \im(\tr(H_{u}^{2}H_{d}H_{u}H_{d}^{3}))=0\\
    & I_{6}= \im(\tr(H_{u}^{3}H_{d}H_{u}H_{d}^{3}))=0\\
    & I_{7}= \im(\tr(H_{u}^{2}H_{d}H_{u}H_{d}^{4}-H_{u}^{2}H_{d}^{2}H_{u}H_{d}^{3}))=0\\
    & I_{8}=
    \im(\tr(H_{u}^{2}H_{d}H_{u}H_{d}^{5}-H_{u}^{2}H_{d}^{2}H_{u}H_{d}^{4}
    +H_{u}H_{d}H_{u}H_{d}^{2}H_{u}H_{d}^{3}))=0.
  \end{aligned}\label{eq:4}
\end{equation}
All invariants in~\eqref{eq:3} and $I_{k}$ in~\eqref{eq:4} can be
written as sums of expressions that contain the function $G(i,j;k,l)$
(where $m_{u_{i}}$ and $m_{d_{i}}$ are the \textit{up} and
\textit{down} quark masses, respectively)
\begin{equation}\label{eq:5}
  G(i,j;k,l)= -(m_{u_{j}}^{2}-m_{u_{i}}^{2})(m_{u_{4}}^{2}-m_{u_{i}}^{2})(m_{u_{4}}^{2}-m_{u_{j}}^{2})
  \times(m_{d_{l}}^{2}-m_{d_{k}}^{2})(m_{d_{4}}^{2}-m_{d_{k}}^{2})(m_{d_{4}}^{2}-m_{d_{l}}^{2})
  \im(V_{ik}V_{jk}^{*}V_{jl}V_{il}^{*}).
\end{equation}
multiplied by polynomials of squares of quark masses. From
Eq.~\eqref{eq:5} one can see that the \textit{Jarlskog} type
invariants also play an important role in the conditions for the
CP~conservation for the SM4.

The CKM matrix in the SM4 has three phases and the vanishing of
these~3~phases is also a condition for CP~conservation. The most
remarkable fact is that there are 8~relations in Eq.~\eqref{eq:4},
which put conditions on the~3~phases. It thus seems that 8~conditions
in Eq.~\eqref{eq:4} contain a mixture of \textit{fine tuned}
conditions and also conditions for the CKM matrix only. The
\textit{fine tuned} conditions imply the vanishing of the phases, so
eventually the conditions for the CKM matrix are important. In this
paper we will find the necessary and sufficient conditions for the CP
conservation in the SM4 in terms of the \textit{Jarlskog} invariants
only.

\section{Rephasing transformations}
\label{sec:reph-transf}

In our study we will consider only the case of non degenerate quark
masses within the up and down quarks. In such a case the SM is
invariant under the rephasing transformation of the quark fields (and
not a bigger unitary group) and this implies that the CKM matrix is
determined up to the rephasing transformation. From this it follows
that the~3~generations CKM matrix has~1~independent phase and
3~angles, while the~4~generation CKM matrix depends on 3~independent
phases and 6~angles. The vanishing of these phases is a sufficient and
necessary condition for the CP~conservation. In general the
CP-conservation means that for $n$-generations there must exist two
diagonal matrices
$D_{L}=\diag(\me^{i\phi_{1}},\me^{i\phi_{2}},\dots,\me^{i\phi_{n}})$
and $D_{R}=\diag(1,\me^{i\psi_{1}},\dots,\me^{i\psi_{n-1}})$, such
that the following equation is fulfilled
\begin{equation}
  \label{eq:6}
  \im(D_{L}V_{\text{CKM}}D_{R})=0.
\end{equation}
One might consider Eq.~\eqref{eq:6} as an analogue of
condition~\eqref{eq:1}, but the mathematical requirements for
Eqs.~\eqref{eq:1} and~\eqref{eq:6} are different since
Eq.~\eqref{eq:6} is applied only to one matrix, $V_{\text{CKM}}$, and
the matrices $D_{L}$ and $D_{R}$, which are to be determined, are
\textit{diagonal} unitary matrices. Step by step we will find the
conditions on the CKM4 matrix, stemming from the CP-conservation.

As a first step we will find conditions for a general $n\times n$
matrix to be equivalent to a real matrix as a result of a rephasing
transformation. These conditions are given by the following
\begin{thm}\label{thm:1}
  A $n\times n$ complex matrix $M=m_{ij}\neq 0$ is equivalent to a
  real $n\times n$ matrix $\widetilde{M}=\vert m_{ij}\vert$ by the
  following rephasing transformation, $M = D_{L}\widetilde{M}D_{R}$,
  iff
  \begin{equation}\label{eq:7}
    \im(m^{\phantom{*}}_{11}m^{\phantom{*}}_{kl}m_{1l}^{*}m_{k1}^{*})=0,\;\; k,l=2,\ldots,n.
  \end{equation}
  Here
  \begin{equation*}
    D_{L}=\diag(\me^{i\phi_{1}},\me^{i\phi_{2}},\dots,\me^{i\phi_{n}}),\quad
    D_{R}=\diag(1,\me^{i\psi_{1}},\dots,\me^{i\psi_{n-1}})
  \end{equation*}
  with $\phi_{i}$ and $\psi_{i}$ being real phases.
\end{thm}
The proof of this theorem is given in the Appendix. From
Theorem~\ref{thm:1} we see that for a $n\times n$ arbitrary matrix
there are $(n-1)^{2}$ conditions for the rephasing equivalence to a
real matrix, i.e., for a $3\times3$ matrix there are~4~conditions and
for a $4\times4$ matrix there are 9~conditions. Also note that the
conditions are expressed in terms of the rephasing invariants of
\textit{Jarlskog} type, $\im(m_{ij}m_{kl}m^{*}_{il}m^{*}_{kj})$,
$i<k$, $j<l$, and do not require calculation of the phases $\phi_{i}$
and $\psi_{i}$. One should also see that there is an additional
freedom in condition~\eqref{eq:7}: instead of $m_{11}$ one can choose
an arbitrary fixed element $m_{i_{0}j_{0}}$ and then
$k\in 1,2,\ldots,n$ and $k\neq i_{0}$ and $l\neq j_{0}$. For a general
$n\times n$ matrix there exist $\Big(\frac{n(n-1)}{2}\Big)^{2}$
\textit{Jarlskog} invariants (see Appendix), and Theorem~\ref{thm:1}
determines that only $(n-1)^{2}$ conditions are necessary for the
remaining \textit{Jarlskog} invariants to vanish. This implies the
existence of relations between \textit{Jarlskog} invariants for $n>2$.

Condition~\eqref{eq:6} for the CP-conservation is imposed on the CKM
matrix, which is \textit{unitary} while Theorem~\ref{thm:1} gives its
equivalence to a real matrix for an arbitrary matrix. The unitarity of
a matrix reduces the number of conditions for the
CP-conservation. This happens, because the unitarity of a matrix (in
particular CKM) imposes relations between the \textit{Jarlskog}
invariants.

The next step in our analysis is to consider the relations between the
\textit{Jarlskog} invariants for a unitary matrix, which are given in
the following
\begin{thm}\label{thm:2}
  The unitarity of the $n\times n$ matrix $V$ implies the following
  set of linear relations between the \textit{Jarlskog} invariants
  \begin{subequations}\label{eq:8}
    \begin{align}
      \label{eq:8a}
      &\im\Big(\sum_{l\neq i}V^{\phantom{*}}_{ij}V^{\phantom{*}}_{lk}V^{*}_{ik}V^{*}_{lj}\Big)=0,\quad j<k, \;\; i,j=1,\ldots,n,\;k=2,\ldots,n\\ \label{eq:8b}
      &\im\Big(\sum_{l\neq i}V^{\phantom{*}}_{ji}V^{\phantom{*}}_{kl}V^{*}_{jl}V^{*}_{ki}\Big)=0,\quad j<k, \;\; i,j=1,\ldots,n,\;k=2,\ldots,n.
    \end{align}
  \end{subequations}
\end{thm}
The proof of the theorem is given in the
Appendix. Relations~\eqref{eq:8a} follow from the orthogonality of the
rows of the matrix $V$ and relations~\eqref{eq:8b} follow from the
orthogonality of the columns. There are altogether $n^{2}(n-1)$
relations of the type~\eqref{eq:8} between
$\Big(\frac{n(n-1)}{2}\Big)^{2}$ \textit{Jarlskog} invariants. Not all
relations~\eqref{eq:8} are linearly independent, in the case $n=3$
remains~1~independent invariant and for $n=4$ remain~4~independent
invariants.

\section{Conditions for the CP conservation}
\label{sec:cond-cp-cons}

\subsection*{General case}
Let us start from the case of the $n\times n$ CKM matrix $V$. From
Theorem~\ref{thm:1} the conditions for the CKM matrix to be real after
the rephasing are
\begin{equation}
  \label{eq:13}
  \im(V^{\phantom{*}}_{11}V^{\phantom{*}}_{ij}V^{*}_{i1}V^{*}_{1j})=0, \quad i,j=2,\ldots,n.
\end{equation}
Eq.~\eqref{eq:13} requires the vanishing of the $(n-1)^{2}$
\textit{Jarlskog} invariants for the CP conservation.

The CKM matrix is unitary and from Theorem~\ref{thm:2} we know that
the \textit{Jarlskog} invariants are linearly dependent and those from
Eq.~\eqref{eq:13} fulfill the relations
\begin{subequations}\label{eq:14}
  \begin{align}
    \label{eq:14a}
    \im\Big(\sum_{l=2}^{n} V^{\phantom{*}}_{11}V^{\phantom{*}}_{lk}V^{*}_{1k}V^{*}_{l1}\Big)=0, \quad k=2,\ldots,n,\\ \label{eq:14b}
    \im\Big(\sum_{l=2}^{n} V^{\phantom{*}}_{11}V^{\phantom{*}}_{kl}V^{*}_{1l}V^{*}_{k1}\Big)=0, \quad k=2,\ldots,n.
  \end{align}
\end{subequations}
Eqs.~\eqref{eq:14} contain $2(n-1)$ relations, which are not linearly
independent and there exist one relation between them: the sum of
Eqs.~\eqref{eq:14a} is equal to the sum of Eqs.~\eqref{eq:14b}. This
means that Eqs.~\eqref{eq:14} reduce the number of conditions by
$2(n-1)-1$. Thus after including the unitarity of the CKM matrix the
number of conditions for the CP conservation is equal
\begin{equation}
  \label{eq:15}
  \underbrace{(n-1)^{2}}_{\text{number of conditions in Eq.~\eqref{eq:13}}}- \underbrace{(2(n-1)-1)}_{\text{number of relations in Eq.~\eqref{eq:14}}} =(n-2)^{2}.
\end{equation}
The number of conditions in Eq.~\eqref{eq:15} should be compared to
the number of phases in the CKM matrix, what is done in
Table~\ref{tab:1}.
\begin{table}[ht]
  \centering
  \begin{tabular}{r|c|c|c}
    number of quark generations&3&4&5\\[3pt] \hline
    \vphantom{{\Large A}}number of phases &1&3&6\\[3pt] \hline
    \vphantom{{\Large A}}number of conditions in Eq.~\eqref{eq:15}&1&4&9
  \end{tabular}
  \caption{Comparison of the number of phases of the CKM matrix and
    the number of conditions in Eq.~\eqref{eq:15}.}
  \label{tab:1}
\end{table}
From Table~\ref{tab:1} one confirms the well known fact that
for~3~generations the number of conditions is equal to the number of
phases. For more generations the number of conditions is larger than
the number of the phases. By the parameter counting these two numbers
should be equal. One thus may ask the question: \textit{Is there a
  mechanism which makes that these two numbers should differ or there
  is a way to prove that these two numbers are equal?} The answer is
in the next subsection.

\subsection*{CP violation for 4 quark generations}

To find conditions for the CP conservation for 4~generations we will
use an explicit, \textit{slightly modified} parameterization of the
CKM4 matrix from~\cite{Perez:arxiv1209.5812,R.:2016xzg}. As we know the CKM4 matrix is
parameterized by 6~angles
$\alpha_{12},\alpha_{13},\alpha_{14},\alpha_{23},\alpha_{24},\alpha_{34}$
and 3~phases $\beta_{13},\beta_{14},\beta_{24}$. We will not write
here the full expressions for the CKM4 matrix, but we will only
include the following three \textit{Jarlskog} invariants
($c_{ij}=\cos\alpha_{ij}$ and $s_{ij}=\sin\alpha_{ij}$)
\begin{subequations}
  \label{eq:16}
  \begin{align}
    \label{eq:16a}
    &\im(V^{\phantom{*}}_{11}V^{\phantom{*}}_{22}V^{*}_{12}V^{*}_{21}) = -\sin\beta_{14} c_{13} c_{14}^2 c_{24} s_{13} s_{14} s_{23} s_{24}\\
    &\im(V^{\phantom{*}}_{11}V^{\phantom{*}}_{23}V^{*}_{13}V^{*}_{21}) = \cos\beta_{14} \sin\beta_{13} c_{12} c_{13} c_{14}^2 c_{23} c_{24} s_{12} s_{14} s_{24}\nonumber\\
    &\phantom{\im(V^{\phantom{*}}_{11}V^{\phantom{*}}_{22}V^{*}_{12}V^{*}_{21})=} +\cos\beta_{13}
      \sin\beta_{14} c_{12} c_{13} c_{14}^2 c_{23} c_{24} s_{12} s_{14} s_{24}\nonumber\\ \label{eq:16b}
    &\phantom{\im(V^{\phantom{*}}_{11}V^{\phantom{*}}_{22}V^{*}_{12}V^{*}_{21})=} +\sin\beta_{14} c_{12}^2 c_{13} c_{14}^2 c_{24}
      s_{13} s_{14} s_{23} s_{24}\\
    &\im(V^{\phantom{*}}_{11}V^{\phantom{*}}_{32}V^{*}_{12}V^{*}_{31}) = -\sin\beta_{24} c_{13} c_{14}^2 c_{23} c_{24} c_{34} s_{13} s_{14} s_{34}\nonumber\\ \label{eq:16c} &\phantom{\im(V^{\phantom{*}}_{11}V^{\phantom{*}}_{32}V^{*}_{12}V^{*}_{31}) =} +\sin\beta_{14}c_{13} c_{14}^2 c_{24} s_{13} s_{14} s_{23} s_{24} s_{34}^2.
  \end{align}
\end{subequations}
From Eqs.~\eqref{eq:16} we obtain:
\begin{align*}
  &\text{From Eq.~\eqref{eq:16a}: } \im(V^{\phantom{*}}_{11}V^{\phantom{*}}_{22}V^{*}_{12}V^{*}_{21}) =0\;\Leftrightarrow
    \beta_{14}=0\text{ or }\pi,\\
  &\text{From Eqs.~\eqref{eq:16a} and~\eqref{eq:16b}: } \im(V^{\phantom{*}}_{11}V^{\phantom{*}}_{22}V^{*}_{12}V^{*}_{21}) =0\text{ and } \im(V^{\phantom{*}}_{11}V^{\phantom{*}}_{23}V^{*}_{13}V^{*}_{21}) =0\\ &\phantom{\text{From Eq.~\eqref{eq:16b}: }}
                                                                                                                                                                                                                  \Leftrightarrow \beta_{14}=0\text{ or }\pi\text{ and }\beta_{13}=0\text{ or }\pi,\\
  &\text{From Eqs.~\eqref{eq:16a} and~\eqref{eq:16c}: } \im(V^{\phantom{*}}_{11}V^{\phantom{*}}_{22}V^{*}_{12}V^{*}_{21}) =0\text{ and } \im(V^{\phantom{*}}_{11}V^{\phantom{*}}_{32}V^{*}_{12}V^{*}_{31}) =0\\ &\phantom{\text{From Eq.~\eqref{eq:16b}: }}
                                                                                                                                                                                                                  \Leftrightarrow
                                                                                                                                                                                                                  \beta_{14}=0\text{ or }\pi\text{ and }\beta_{24}=0\text{ or }\pi.
\end{align*}
We thus see that we have obtained
\begin{thm}\label{thm:3}
  \begin{equation*}
    \left(
      \begin{array}{l}
        \im(V^{\phantom{*}}_{11}V^{\phantom{*}}_{22}V^{*}_{12}V^{*}_{21}) =0\\
        \im(V^{\phantom{*}}_{11}V^{\phantom{*}}_{23}V^{*}_{13}V^{*}_{21}) =0\\
        \im(V^{\phantom{*}}_{11}V^{\phantom{*}}_{32}V^{*}_{12}V^{*}_{31}) =0
      \end{array}\right) \Leftrightarrow
    (\beta_{13}=0,\;\beta_{14}=0,\;\beta_{24}=0) .
  \end{equation*}
\end{thm}
The vanishing of the phases $\beta_{ij}$ of the CKM4 means that CP is
conserved.

We have thus obtained the condition for the CP conservation in
SM4. From Theorem~\ref{thm:3} we obtain the conditions for the CP
violation in SM4
\begin{thm}\label{thm:4} {\rm (symbol } $\vee$ {\rm denotes the
    logical} \textit{or}{\rm )}
  \begin{equation*}
    (\im(V^{\phantom{*}}_{11}V^{\phantom{*}}_{22}V^{*}_{12}V^{*}_{21})
    \neq0 \vee
    \im(V^{\phantom{*}}_{11}V^{\phantom{*}}_{23}V^{*}_{13}V^{*}_{21})
    \neq0 \vee
    \im(V^{\phantom{*}}_{11}V^{\phantom{*}}_{32}V^{*}_{12}V^{*}_{31}) \neq0)
    \Leftrightarrow (\beta_{13}\neq0
    \vee\beta_{14}\neq0\vee\beta_{24}\neq0).
  \end{equation*}
\end{thm}
It means that the non vanishing of any of the \textit{Jarlskog}
invariant implies the presence of the CP violation in SM4.

\section{Discussion and conclusions}
\label{sec:disc-concl}

Theorems~\ref{thm:3} and~\ref{thm:4} constitute the main results of
the paper and are the generalization of the Jarlskog's conditions for
CP conservation is the SM with 3~generations. Let us comment on these
results.
\begin{enumerate}
\item Theorem~\ref{thm:4} states that the non vanishing of any
  \textit{Jarlskog} invariant implies the CP violation. Since CP is
  broken in SM3, then it means that is violated in SM4 also.
\item Theorems~\ref{thm:3} and~\ref{thm:4} put conditions on the three
  following \textit{Jarlskog} invariants:
  $\im(V^{\phantom{*}}_{11}
  V^{\phantom{*}}_{22}V^{*}_{12}V^{*}_{21})$,
  $ \im(V^{\phantom{*}}_{11}
  V^{\phantom{*}}_{23}V^{*}_{13}V^{*}_{21})$,
  $\im(V^{\phantom{*}}_{11}
  V^{\phantom{*}}_{32}V^{*}_{12}V^{*}_{31})$. These three invariants
  are built from the $3\times3$ CKM matrix alone. Had the CP been
  conserved in the SM4, then it would be possible to verify it from
  the CKM matrix of the SM3.
\item In Theorems~\ref{thm:3} and~\ref{thm:4} we can use any three
  \textit{Jarlskog} invariants
  $\im(V^{\phantom{*}}_{ij}
  V^{\phantom{*}}_{kl}V^{*}_{il}V^{*}_{kj})$, but the indices
  $(i,j,k)$ or $(i,j,l)$ of all of these three invariants cannot be
  equal.
\item The unitarity conditions~\eqref{eq:8} of the CKM matrix written
  for the CKM4 matrix have the form
  \begin{subequations}\label{eq:23}
    \begin{align}
      \label{eq:23a}
      \im(V^{\phantom{*}}_{11} V^{\phantom{*}}_{22}V^{*}_{12}V^{*}_{21}) +\im(V^{\phantom{*}}_{11} V^{\phantom{*}}_{32}V^{*}_{12}V^{*}_{31}) +\im(V^{\phantom{*}}_{11} V^{\phantom{*}}_{42}V^{*}_{12}V^{*}_{41})=0\\
      \intertext{and}
      \label{eq:23b}
      \im(V^{\phantom{*}}_{11} V^{\phantom{*}}_{22}V^{*}_{12}V^{*}_{21}) +\im(V^{\phantom{*}}_{11} V^{\phantom{*}}_{23}V^{*}_{13}V^{*}_{21}) +\im(V^{\phantom{*}}_{11} V^{\phantom{*}}_{24}V^{*}_{14}V^{*}_{21})=0.
    \end{align}
  \end{subequations}
\end{enumerate}
The measurement of the first two \textit{Jarlskog} invariants in each
of equations~\eqref{eq:23} does not involve particles from the 4-th
generation. The following conditions on these invariants
\begin{subequations}\label{eq:24}
  \begin{align}
    \label{eq:24a}
    \im(V^{\phantom{*}}_{11} V^{\phantom{*}}_{22}V^{*}_{12}V^{*}_{21}) +\im(V^{\phantom{*}}_{11} V^{\phantom{*}}_{32}V^{*}_{12}V^{*}_{31})\neq 0\\
    \intertext{or}
    \label{eq:24b}
    \im(V^{\phantom{*}}_{11} V^{\phantom{*}}_{22}V^{*}_{12}V^{*}_{21}) +\im(V^{\phantom{*}}_{11} V^{\phantom{*}}_{23}V^{*}_{13}V^{*}_{21})\neq 0
  \end{align}
\end{subequations}
would imply the existence of the 4-th quark generation or non
unitarity of the CKM3 matrix.

\appendix
\setcounter{secnumdepth}{0}
\renewcommand{\theequation}{A.\arabic{equation}}
\setcounter{equation}{0}

\section{Appendix}
The \textit{Jarlskog} invariant of the CKM matrix is defined by the
following expression
\begin{equation}
  \label{eq:11}
  \im(V^{\phantom{*}}_{ij}V^{\phantom{*}}_{kl}V^{*}_{il}V^{*}_{kj}),\quad i\neq k\text{ and }j\neq l
\end{equation}
and it is invariant under the rephasing transformation. It fulfills
two simple identities
\begin{equation}
  \label{eq:12}
  \im(V^{\phantom{*}}_{ij}V^{\phantom{*}}_{kl}V^{*}_{il}V^{*}_{kj}) = \im(V^{\phantom{*}}_{kl}V^{\phantom{*}}_{ij}V^{*}_{kj}V^{*}_{il}) = -\im(V^{\phantom{*}}_{il}V^{\phantom{*}}_{kj}V^{*}_{ij}V^{*}_{kl}).
\end{equation}
It means that it is always possible to rearrange the order of indices
in such a way that $i<k$ and $j<l$. Taking this into account it is
simple to demonstrate that for the $n\times n$ matrix there are
$\Big(\frac{n(n-1)}{2}\Big)^{2}$ independent \textit{Jarlskog}
invariants.
\subsection{Proof of Theorem~\ref{thm:1}}
\label{proof1}
\noindent $\Rightarrow$ From $M = D_{L}\widetilde{M}D_{R}$,
($\widetilde{M}=\vert m_{ij}\vert$), we have
\begin{equation}
  \label{eq:17}
  \begin{aligned}
    & m_{11}=\me^{i\phi_{1}}\vert m_{11}\vert,\;
    m_{1l}=\me^{i\phi_{1}}\me^{i\psi_{l-1}}\vert m_{1l}\vert,\\
    & m_{k1}=\me^{i\phi_{k}}\vert m_{k1}\vert,\;
    m_{kl}=\me^{i\phi_{k}}\me^{i\psi_{l-1}}\vert m_{kl}\vert,
  \end{aligned}\quad k,l>1.
\end{equation}
From Eq.~\eqref{eq:17} we have
\begin{equation}
  \label{eq:18}
  \im(m^{\phantom{*}}_{11}m^{\phantom{*}}_{kl}m_{1l}^{*}m_{k1}^{*})
  = \im(\me^{i\phi_{1}}\vert m_{11}\vert
  \me^{i\phi_{k}}\me^{i\psi_{l-1}}\vert m_{kl}\vert
  \me^{-i\phi_{1}}\me^{-i\psi_{l-1}}\vert m_{1l}\vert
  \me^{-i\phi_{k}}\vert m_{k1}\vert)
  = \im(\vert m_{11}\vert \vert m_{kl}\vert \vert m_{1l}\vert \vert
  m_{k1}\vert)=0.
\end{equation}

\noindent $\Leftarrow$ Now, when
$m_{ij}=\vert m_{ij}\vert \me^{i\omega_{ij}}$. The condition
$\im(m^{\phantom{*}}_{11}m^{\phantom{*}}_{kl}m_{1l}^{*}m_{k1}^{*})=0$
implies
\begin{equation}
  \label{eq:19}
  \omega_{11}+\omega_{kl}-\omega_{1l}-\omega_{k1}=0\text{ or }\pi.
\end{equation}
Let us define
\begin{equation}
  \label{eq:20}
  \phi_{k}=\omega_{k1},\; \psi_{l-1}=\omega_{1l}-\omega_{11}.
\end{equation}
From Eq.~\eqref{eq:19} we find
\begin{equation}
  \label{eq:21}
  \phi_{k}+\psi_{l-1}=\omega_{k1}+\omega_{1l}-\omega_{11}=\omega_{kl},
\end{equation}
thus
$m_{kl}=\me^{i\omega_{kl}}\vert
m_{kl}\vert=\me^{i\phi_{k}}\me^{i\psi_{l-1}} \vert
m_{kl}\vert$. Denoting
$D_{L}= \diag(\me^{i\phi_{1}},\me^{i\phi_{2}},\dots,\me^{i\phi_{n}})$
and $D_{R}=\diag(1,\me^{i\psi_{1}},\dots,\me^{i\psi_{n-1}})$ we obtain
\begin{equation}
  \label{eq:22}
  M = D_{L}\widetilde{M}D_{R}
\end{equation}
and this complete the proof.
\subsection{Proof of Theorem~\ref{thm:2}}
\label{proof2}

The CKM matrix is unitary, which imposes unitary relations between its
matrix elements. Such relations also imply additional relations
between the \textit{Jarlskog} invariants. The standard method of the
derivation of such relations is to multiply the $i$-th column (or row)
by the compex conjugate of the $j$-th column (or row) of the
$n\times n$ unitary matrix $V$
\begin{equation}
  \label{eq:9}
  \sum_{l=1}^{n}V^{\phantom{*}}_{lk}V^{*}_{lj}=0,\text{ for }j\neq k.
\end{equation}
\begin{subequations}\label{eq:10}
  Now, if we multiply Eq.~\eqref{eq:9} by
  $V^{\phantom{*}}_{ij}V^{*}_{ik}$ we obtain
  \begin{equation}
    \label{eq:10a}
    \sum_{l\neq i} \im(V^{\phantom{*}}_{ij}V^{\phantom{*}}_{lk}V^{*}_{ik}V^{*}_{lj})=0,\;\; j<k,\quad i,j,k=1,\ldots,n.
  \end{equation}
  Multiplying rows we obtain in the same way
  \begin{equation}
    \label{eq:10b}
    \sum_{l\neq i} \im(V^{\phantom{*}}_{ji}V^{\phantom{*}}_{kl}V^{*}_{jl}V^{*}_{ki})=0,\;\; j<k,\quad i,j,k=1,\ldots,n
  \end{equation}
\end{subequations}
and this completes the proof.

\end{document}